\documentclass[sigconf,screen,review=false]{acmart}

\usepackage{algorithm}
\usepackage{algorithmic}
\usepackage{appendix}
\usepackage{array}
\usepackage{balance}
\usepackage{booktabs}
\usepackage{color}
\usepackage{comment}
\usepackage{datetime}
\usepackage{enumerate}
\usepackage{enumitem}
\usepackage{fancyhdr}
\usepackage{framed}
\usepackage{graphicx}
\usepackage{hyperref}
\usepackage{listings}
\usepackage{microtype}
\usepackage{multirow}
\usepackage{soul}
\usepackage{subfig}
\usepackage{textcomp}
\usepackage{textgreek}
\usepackage{url}
\usepackage{xcolor}
\usepackage{xspace}
\usepackage{cleveref}

\def\BibTeX{{\rm B\kern-.05em{\sc i\kern-.025em b}\kern-.08em
    T\kern-.1667em\lower.7ex\hbox{E}\kern-.125emX}}

\newcommand{\ignore}[1]{}

\copyrightyear{2020}
\acmYear{2020}
\setcopyright{acmlicensed}\acmConference[HASP '20]{Hardware and Architectural Support for Security and Privacy}{October 17, 2020}{Virtual, Greece}
\acmBooktitle{Hardware and Architectural Support for Security and Privacy (HASP '20), October 17, 2020, Virtual, Greece}
\acmPrice{15.00}
\acmDOI{10.1145/3458903.3458911}
\acmISBN{978-1-4503-8898-6/20/10}

\title{SoK: Opportunities for Software-Hardware-Security Codesign for Next Generation Secure Computing}
\author{Deeksha Dangwal$^1$, Meghan Cowan$^1$, Armin Alaghi$^1$, Vincent T. Lee$^1$, Brandon Reagen$^2$,
  Caroline Trippel$^3$}
\email{{ddangwal, meghancowan, alaghi, vtlee}@fb.com,
bjr5@nyu.edu, trippel@stanford.edu}
\affiliation{
  $^1$ Facebook Reality Labs Research \\
  $^2$ New York University \\
  $^3$ Stanford University}

\begin{document}
\begin{abstract}

Users are demanding increased data security.
As a result, security is rapidly becoming a first-order design constraint in next generation computing systems.
Researchers and practitioners are exploring various security technologies to meet user demand such as trusted execution environments (e.g., Intel SGX, ARM TrustZone), homomorphic encryption, and differential privacy.
Each technique provides some degree of security, but differs with respect to threat coverage, performance overheads, as well as implementation and deployment challenges.
In this paper, we present a systemization of knowledge (SoK) on these design considerations and trade-offs using several prominent security technologies.
Our study exposes the need for \textit{software-hardware-security} codesign to realize efficient and effective solutions of securing user data.
In particular, we explore how design considerations across applications, hardware, and security mechanisms must be combined to overcome fundamental limitations in current technologies so that we can minimize performance overhead while achieving sufficient threat model coverage.
Finally, we propose a set of guidelines to facilitate putting these secure computing technologies into practice.

\end{abstract}

\maketitle
\thispagestyle{firstpage}
\pagestyle{plain}
\sloppy
\setlength{\parskip}{0.1cm plus2mm minus2mm}


\section{Introduction}
\label{sec:introduction}

Over the past decade, data security has emerged as a first-order design constraint.
Users have begun to demand increased accountability from data aggregators; they want to know how their data is being managed and protected against misuse or abuse.
As a result, the development of and interest in secure computing technologies has progressed rapidly over the last few years.
For example, technologies such as trusted execution environments (TEEs), homomorphic encryption (HE), and differential privacy (DP) are all rapidly maturing areas of research.

Despite recent advances, each technology presents a range of trade-offs and challenges; these challenges include:
(1) implementing the secure computing technology,
(2) verifying that the security guards are correctly implemented and provide intended threat model support,
(3) porting the technology to support a variety of applications,
and (4) maintaining practical performance targets.

Secure computing solutions present trade-offs that impact security and performance which must be carefully balanced to satisfy performance and security specification targets.
We argue that, if carefully engineered, secure computing technologies can enable the best of both worlds (good performance and security) instead of mandating one or the other (performance or security).
For example, TEEs offer confidentiality and integrity for computations executing within an isolated memory location, i.e., an enclave.
TEE support for an application like machine learning (ML) inference on sensitive data would be critical.
To deploy the application, the designer needs to port the ML inference to work with the chosen TEE's API and consider the cost of each call into the TEE.
For example, SGX enclaves have limited memory capacity.
Therefore large ML models need to be encrypted and stored on a larger, enclave-external memory and the resulting computation partitioned properly.
Model partitions must be moved into the SGX enclave, which causes page swapping overheads.
Combined, these design considerations require expertise across the software/application, hardware, and security in tandem to achieve an efficient and secure implementation.
Overcoming these challenges mandates a careful re-examination of complex cross-stack design choices to find effective and efficient solutions.

More broadly, software-hardware-security codesign has emerged as a design philosophy for realizing performance- and power-efficient secure computing technologies.
This notion has historical precedence; software-hardware codesign has a proven track record in improving the overall efficiency of computing solutions that we build to support modern applications.
Examples include fixed-point approximation for neural networks, domain-specific accelerators like digital signal processor (DSP) units, and hardware-friendly mathematical approximations.
These examples of software-hardware codesign have had tremendous practical impact; aggressive fixed-point quantization has enabled power-efficient implementations of machine learning and DSPs have enabled high performance telecommunication applications that form the backbone of mobile networks.
However, these innovations required feedback between software and hardware design considerations to arrive at the highly optimized solutions and architectures we have today.

A valuable codesign process by adding security to the mix can analogously enhance the efficacy of secure computing technologies.
To address the added complexities introduced by security considerations, experts from across hardware, software, and security need to be able to communicate through systematic and well-defined abstraction boundaries.
Unfortunately, the process of identifying these opportunities is not well systemized; many seemingly disparate computing techniques fall into the category of codesign.
It is, therefore, valuable to concretely systemize when there are codesign opportunities to allow this design iteration to occur for emerging secure computing technologies such as TEEs, HE, and DP.

{
\setlength{\belowcaptionskip}{-2mm}
\begin{figure}[t]
    \centering
    \includegraphics[width=\linewidth]{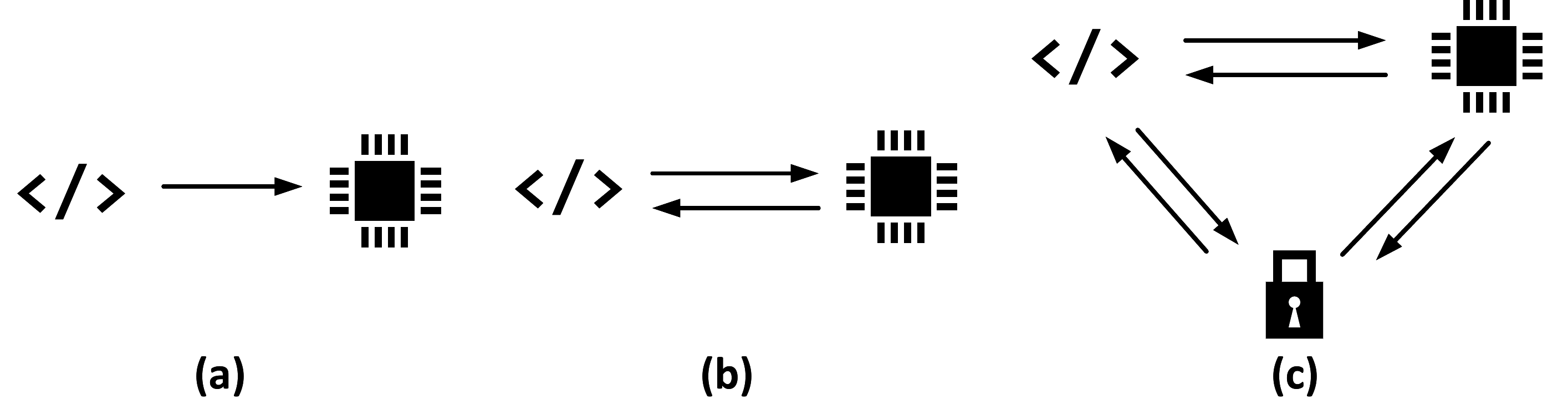}
    \caption{Hardware acceleration versus software-hardware codesign and software-hardware-security codesign. (a) Hardware acceleration does not integrate hardware design feedback. (b) Software/hardware codesign integrates feedback. (c) Software-hardware-security codesign additionally integrates security considerations.
    }
    \label{fig:sw_hw_codesign}
\end{figure}
}

We propose a systemization of knowledge to analyze the hardware, software, and security design considerations required to realize practical and efficient solutions for emerging secure computing technologies.
We start by defining the notion of software-hardware-security codesign (\autoref{sec:background}).
We then show there is historical precedence in the selection process of the advanced encryption standard and apply our systemization to emerging secure computing technologies (\autoref{sec:systemization}): differential privacy, trusted execution environments, and homomorphic encryption.
We find that a common theme across these secure computing technologies is the prevalence of \textit{feedback loops} to inform and drive the implementations towards better designs.
Finally, we synthesize these insights (\autoref{sec:themes}) to motivate the need to put software-hardware-security codesign into practice and enable security as a first-order design constraint.

\section{Codesign Methodologies}
\label{sec:background}

This section defines the notion of software-hardware-security codesign and shows it is a natural extension to the existing software-hardware codesign process.

\subsection{Software-Hardware Codesign}

Software-hardware codesign is a popular concept that originated in the 1990s as a way to co-optimize embedded systems software with the underlying hardware implementation~\cite{kalavade1993hardware}.
Software-hardware codesign \textit{simultaneously considers both hardware and software constraints to make better design decisions and improve the overall quality of results (i.e., performance, power, accuracy, etc.) of the system} (\autoref{fig:sw_hw_codesign}b).
The key distinction between standard \textit{hardware acceleration} (\autoref{fig:sw_hw_codesign}a) and \textit{software-hardware codesign} is an iterative feedback loop between software and hardware considerations.
Hardware acceleration is a one-way process where the application is not affected by hardware constraints; software-hardware codesign allows design consideration feedback in either direction between software and hardware.
This distinction is important because this feedback loop is what enables co-optimization of the software and hardware to ultimately yield a more efficient overall solution.

A canonical example of software-hardware codesign when building a hardware accelerator is fixed-point approximation (i.e., fixed-point optimization).
In fixed-point approximation, a designer may convert over-provisioned floating-point units to fixed-point arithmetic since hardware implementations of floating points require more power, area, and latency.
However, replacing floating-point with fixed-point units introduces some error to the original functionality.
For this substitution to be acceptable, the designer must revisit the application to ensure that the approximation does not adversely impact accuracy metrics.
In this case, the application drives the initial hardware accelerator design but power-efficiency considerations feeds information back to the application design.
The impact on accuracy and performance further informs whether the quantization approximation is acceptable; if not, the hardware is adjusted to provide better fidelity and the process iterates.

\subsection{Software-Hardware-Security Codesign}

It is natural to extend this idea of having a tightly coupled, synergistic codesign feedback loop to include security constraints.
We define the notion of software-hardware-security codesign as \emph{a design process which simultaneously considers software, hardware, and security design parameters when optimizing for power- and performance-efficient, high-fidelity, and threat-model-optimal execution solutions.} (\autoref{fig:sw_hw_codesign}c).
More precisely, we define \textit{software} to encompass both the software implementation (i.e., source code) as well as application design choices.
\textit{Hardware} includes architectural, microarchitectural, and silicon implementation decisions such as compute units, memory sizes, and communication overheads.
\textit{Security} refers to parameters associated with, and constraints imposed by the secure computing technology; for instance, these can manifest as encryption scheme parameters and instruction set restrictions.
We also include threat model considerations as part of the security specifications that each technology tries to mitigate against.

The concept of software-hardware-security codesign has been proposed in the past under different variations such as software-hardware codesign for instances of secure computing technologies.
For example, software-hardware codesign to enhance security has been proposed for resource restricted embedded systems such as ~\cite{vrased} for remote attestation and for implementing minimal roots of trust~\cite{eldefrawy2012smart}.
Power, energy, performance, and area constrained embedded systems must now also consider and satisfy security targets.
Our work seeks to highlight this repeating theme and shed light on the fact that these codesign opportunities should be seriously considered for maturing secure computing technologies.

We organize design consideration insights into a dependency graph to understand the codesign interactions between hardware, software, and security (\autoref{fig:sw_hw_codesign}c).
This serves as our basic blueprint to systemize the codesign considerations between hardware, software, and security.
Each node in the dependency graph represents a collection of design parameters associated with hardware, software, and security.
An edge from \textit{A} to \textit{B}, indicates that changing a design parameter in \textit{A} impacts a design consideration in \textit{B}.

Our goal is to establish a set of design consideration dependencies for each secure computing technology, and expose the feedback loops between the nodes and the opportunities for enabling iterative codesign.
A complete systemization for a secure computing technology will require many edges between nodes to capture every possible interaction.
Similar to how software-hardware codesign delicately balances power, performance, accuracy, area, and energy  costs to meet respective specification targets, security considerations must also be carefully balanced with other implementation costs.
We are \textit{not} proposing sacrificing security for the sake of prioritizing other quality metrics.
Rather we are noting that like power, performance, energy, and area that there is a similar balancing act that can trade-off the strength of the secure computing technology \textit{while still achieving a sufficient security guarantee} that defends against the target threat model.
For instance, protecting against \textit{all} possible threats is typically not feasible; security targets needs to be adjusted to a sufficient level of security to address practical attack scenarios.

Next, we look at instances of software-hardware-security codesign and explore historical precedence before systemizing modern technologies.
By exposing the role of software-hardware-codesign, we expect that the insight will provide guidance into the research directions required to drive future codesign opportunities.

\section{Systemization of Secure Computing Technologies}
\label{sec:systemization}

\begin{table*}[]
\caption{Systemization of software-hardware-security codesign considerations for advanced encryption standard (AES), differential privacy (DP), trusted execution environments (TEEs), and homomorphic encryption (HE). Feedback between software, hardware, and security design considerations highlight the need for codesign for each secure computing technology.}
\label{tab:codesign_taxonomy}
    \begin{tabular}{@{}p{0.03\linewidth}p{0.3\linewidth}p{0.3\linewidth}p{0.3\linewidth}@{}}
    \toprule
        & Software $\leftrightarrows$ Hardware & Software $\leftrightarrows$ Security & Hardware $\leftrightarrows$ Security \\ \midrule
    AES &
      \begin{tabular}[c]{@{}l@{}} $\leftarrow$ Resource and performance results  \\ drives application selection \\ $\rightarrow$ Application selection determines \\ performance and resource needs\end{tabular} &
      \begin{tabular}[c]{@{}l@{}} $\leftarrow$ Security needs and specification drive \\ application design\\ $\leftarrow$ Cryptanalysis determines application \\ security strength\end{tabular} &
      \begin{tabular}[c]{@{}l@{}} $\leftarrow$ Encryption key size impacts hardware \\ efficiency \\ $\rightarrow$ Hardware implementation determines \\ timing and power side-channels\end{tabular} \\ \midrule
      DP &
      \begin{tabular}[c]{@{}l@{}} $\leftarrow$ Hardware support can facilitate or\\ enable DP in software\\ $\rightarrow$ Software drives correlated behavior\\ in hardware and lower levels of the \\ stack\end{tabular} &
      \begin{tabular}[c]{@{}l@{}} $\leftarrow$ DP requires integration of noise and\\ random decisions into application\\ $\rightarrow$ Application selection and behavior\\ defines DP privacy budget\end{tabular} &
      \begin{tabular}[c]{@{}l@{}} $\rightarrow$  Correlated hardware behavior \\ may not be DP and expose side channels\end{tabular} \\ \midrule
      TEE &
      \begin{tabular}[c]{@{}l@{}}$\leftarrow$ Exposes hardware security \\ primitives through ISA extensions\\ $\rightarrow$ Program optimized to leverage \\ hardware ISA extensions\end{tabular} &
      \begin{tabular}[c]{@{}l@{}}$\leftarrow$ Requires application modifications \\ to leverage security primitives\\ $\rightarrow$ Expose application-level threat \\ model requirements\end{tabular} &
      \begin{tabular}[c]{@{}l@{}}$\leftarrow$ Offload root of trust to hardware \\ support\\ $\rightarrow$ Hardware-based remote attestation to \\ enable root of trust\\ $\rightarrow$ Influences security interface to \\ leverage hardware\end{tabular} \\ \midrule
      HE &
      \begin{tabular}[c]{@{}l@{}} $\leftarrow$  Hardware architecture and speedups \\ depend on application parallelism\\ $\rightarrow$ High performance overheads mandates\\ hardware support and acceleration\end{tabular} &
      \begin{tabular}[c]{@{}l@{}} $\leftarrow$ HE constrains set of efficient\\ instructions and levels of logic\\  $\leftarrow$ Requires application quantization to \\ encodable HE values\\ $\rightarrow$ Application behavior determines \\ noise budget and encryption parameter \\ settings\end{tabular} &
      \begin{tabular}[c]{@{}l@{}} $\leftarrow$ Encryption parameter settings \\ determine datapath width and storage \\ size\\  $\leftarrow$ HE scheme determines useful \\ computational primitives for acceleration\end{tabular} \\ \bottomrule
    \end{tabular}
    \end{table*}

This section applies our systemization to three secure computing technologies and exposes the software-hardware-security codesign opportunities.
But first, we start with historical precedence to understand how the codesign process has manifested in the past and then look at emerging secure computing technologies.

\subsection{Historical Precedence: Design of the Advanced Encryption Standard}

The concept of software-hardware-security codesign is not without precedence and has been used in the past to build some of the hardware systems we use today.
A quintessential historical case study is the development and selection process of the Advanced Encryption Standard (AES) which is widely used and available in modern computing stacks.
The development of the AES dates back to 1997 and was motivated by the need for stronger end-to-end encryption protocols~\cite{aes_cfp}.
The algorithm for backing what we know as AES today originally consisted of several finalist proposals: MARS~\cite{mars}, RC6~\cite{rc6}, Rijndael~\cite{rijndael}, Serpent~\cite{serpent}, and Twofish~\cite{twofish}.
Ultimately, the NIST selected the Rijndael algorithm but the process through which they arrived at this conclusion provides valuable insights into the importance of the codesign process.

The algorithm that backs the AES was selected after three rounds of public comment based on software-hardware-security codesign considerations.
The selection process report~\cite{aes_selection} outlines the security, hardware, and software design considerations that were used to ultimately select the final standard.
The selection criteria was split into "1) Security, 2) Cost, and 3) Algorithm and Implementation Characteristics"; the latter two considerations manifested in the public comment periods where studies of the software and hardware implementation characteristics for the proposed algorithms were explicitly solicited.

The AES selection process as a software-hardware-security codesign exercise has many aspects summarized in \autoref{tab:codesign_taxonomy}.
The algorithm selection process was driven by both security considerations as well as software and hardware implementation costs.
Cryptanalysis of the security strength of the proposed algorithms contributed to but did not solely drive the selection of the final algorithm.
Over three years, hardware and software performance analyses were conducted to complement the security analyses; multiple case studies compared the resource efficiency as well as performance of these applications on CPU~\cite{aes_cpu_comparison}, FPGA~\cite{aes_fpga_comparison}, and ASIC platforms~\cite{aes_asic_comparison}.
The selection process also conducted a thorough analysis of aspects such as power and timing side-channel attacks when considering the security strength and viability of the proposed algorithms.
Finally, the process included architectural considerations such as impact of machine word size on security parameters like key size, as well as the impact of restricted computing environments (e.g., systems with limited memory) on the candidate encryption algorithms.

The selection process also contemplated a number of other considerations that go beyond software, hardware, and security.
These included simplicity of the application solution, flexibility, and intellectual property restrictions.
The selection process also examined a range of future-proofing measures such as larger machine word sizes (64-bit machines were not around then), whether to standardize two algorithms in case one was compromised, and ancestry of the algorithms to minimize risk of hidden backdoors.

The impact of the AES standardization and codesign efforts is clearly visible today where it is ubiquitous for securing communication lines and data assets.
Modern implementations of AES now have dedicated hardware acceleration units and instruction set customizations to make them highly optimized.
As a result, AES is a classical case study of how software-hardware-security codesign ultimately led to an efficient and secure computing solution.

\subsection{Differential Privacy}

Differential privacy (DP) is useful when one wants to answer questions about or analyze data in aggregate while protecting individual pieces of data.
DP~\cite{dwork2008differential} is originally a data information retrieval concept that was proposed to protect the privacy of individual entries in an aggregate database.
In a differentially private system, information about individuals is aggregated in a way that does not reveal information any individuals who contributed data to the database while still allowing some public information about the group as a whole.
More generally, we say DP is upheld when the observer of the computation output on the aggregated database does not reveal any individual entry.

DP is typically implemented by adding noise to data assets.
For example, canonical implementations use Laplace mechanisms~\cite{dwork2006calibrating, sarathy2011evaluating}, median mechanisms~\cite{roth2009median}, exponential mechanisms~\cite{mcsherry2007mechanism}, or randomized response mechanisms~\cite{erlingsson2014rappor}.
The addition of noise allows individual data entries to be protected from an adversary.
DP has two variants: local and global.
In global DP, a trusted central database aggregates individual data assets; in the local variant, noise is first added to individual data assets before communicating them to a potentially untrusted central database aggregator.
Local differential privacy (LDP) is more desirable in modern systems because it protects client data before it enters any potentially untrusted system.

For example, LDP is useful when collecting private sensor information from embedded Internet of Things (IoT) systems.
These embedded devices often run on limited compute and power resources.
The guarantees of differentially private algorithms change under such limited resources.
Prior work~\cite{choi2018guaranteeing} for LDP has shown that low-resolution and fixed-point implementations that are prevalent in IoT devices are counterproductive to and hinder private computation.
The authors present new resampling and thresholding techniques that they implement in hardware to continue to provide LDP privacy budget guarantees.
More importantly, this case study illustrates the need to consider hardware restrictions on the effective implementation of an LDP application.

DP also often requires modifying the application to integrate noise mechanisms.
However, DP considerations are not restricted to just the application behavior and applies more broadly to the remainder of the computing stack.
For instance, recent work has proposed differentially private type systems~\cite{reed2010distance, gaboardi2013linear, near2019duet} and programming language support~\cite{barthe2016programming, mcsherry2009privacy, roy2010airavat} which integrates differential privacy mechanisms into the compilation process.
There is also emerging work that exposes the need for DP for memory access patterns~\cite{chan2019foundations, kellaris2017accessing} when outsourcing private applications to an untrusted cloud platform to protect from side-channel attacks, even when running the application within a TEE.

DP is another opportunity for software-hardware-security codesign summarized in \autoref{tab:codesign_taxonomy}.
DP requires adjustment of the application to integrate the noise and random decisions to make the application private.
The application behavior in turn determines the privacy budget and configuration settings for DP to guarantee application privacy.
While an application implementation may be differentially private, the observable behavior across the remainder of the computing stack down to hardware may not be differentially private.
Thus, it is important to also consider hardware systems (ex., memory access) when determining whether the DP security guarantees are enforced as the presence of side channel or other vulnerabilities can defeat the efficacy of software-only DP.
Finally, recent work such as ~\cite{choi2018guaranteeing, maycock2015hardware} also shows that there is room for integrating or enhancing existing hardware support to better facilitate DP software such with thresholding and resampling techniques.

\subsection{Trusted Execution Environments}

Trusted execution environments (TEEs) serve as a proxy for encrypted computation.
Intuitively, TEEs provide an execution environment on a host processor to execute private computation\footnote{Some TEEs also guarantee integrity of the computation} that is protected from observation by others parties on the host processor.
For instance, a client system may request a TEE to set up and offload a private computation to the host processor via remote attestation.
TEEs operate by allocating an isolated hardware-protected memory region called an enclave where a protected application's code and data can reside.
Data within enclave memory can only be accessed by code that also resides within enclave memory and special instructions are provided for invoking enclave code.

Computation within an enclave is protected from inspection by other users and system software (with the exception of side-channel attacks).
Users can leverage the TEE's secure remote attestation protocol to ensure that their data is indeed being executed within the intended TEE and that the encrypted data that is sent to the server.
Memory management features also ensure that the host system cannot tamper with or observe the computation on the supplied data.
With TEEs, rather than placing trust solely in a cryptographic algorithm (as in multi-party computation or homomorphic encryption), clients place trust in the manufacturer's (Intel~\cite{mckeen2013innovative}, ARM~\cite{alves2004trustzone}, AMD~\cite{kaplan2016amd}, etc.) hardware to ensure that their private computations are only ever observed by the TEE compute units (i.e., not the unprotected host processors).

In TEEs, hardware considerations consist of instruction set extensions - a standard architecture software-hardware codesign technique - for facilitating secure remote attestation and secure entry to and exit from enclave code.
This ISA interface (e.g., via SGX ISA extensions) enables software to articulate security requirements at the function granularity.
These ISA extensions enable users to run applications on remote machines with confidentiality and in some cases integrity guarantees with respect to a third party adversary on the host machine.
In particular, these extensions (1) enable a user to send code to execute on a remote machine as an enclave application, (2) verify that their desired code is what is running on the remote TEE, and (3) establish a secure communication channel with the remote TEE to send secret data.
The application in exchange for programming against the ISA interface receives confidentiality and possibly integrity guarantees (for data) with the exception of side channels and integrity guarantees (for code and data) via hardware support for properly partitioned programs. \footnote{Some TEEs (e.g., MIT Sanctum~\cite{costan2016sanctum}) also consider side channels threats while other TEEs (e.g., Intel SGX) do not.}
In other words, software is responsible for providing a partitioning that properly leverages the TEE interface.

TEEs can impose restrictions and modifications to the software or application to fully leverage the advantages while minimizing performance overheads.
The application developer (or in compiler infrastructure) translates an application and its associated security requirements to leverage the software-exposed ISA security interface.
This requires refactoring sensitive application function calls which must be protected to interface with the enclave by annotating valid enclave entry and exit points.
This partitioning requires the application designer to be aware of and minimize the overheads associated with moving data into and out of the enclave.
Correctly partitioned software allows for preserving integrity (e.g., in the case of SGX) for code and data; it also enables data confidentiality guarantees (sometimes with the exception of side channels).
Unfortunately, such a partitioning of the code is currently left up to the user and is not guaranteed to produce a secure result~\cite{lind2017glamdring, SCONE, graphene}.

Given an application, fully offloading the entire computation to a TEE in Intel's SGX implementation can be expensive in terms of performance.
As a result, there have been some proposals to partially secure an application computing pipeline to only incur the overheads for a smaller fraction of the computation and execute the remaining computation in plaintext~\cite{ryffel2019partially, narra2019privacy}.
The high level intuition is that partially encrypted computation will yield intermediary results (unencrypted) which are less sensitive or meaningful.
However, there are no systematic methodologies for determining how to partially partition an application while still satisfying cryptographically strong guarantees which makes this an iterative process to balance performance and security levels.
Researchers are also exploring ways to produce such a portion automatically~\cite{lind2017glamdring}.

Combined, these design considerations illustrate how TEEs are another example of how software, hardware, and security must be co-optimized together as summarized in \autoref{tab:codesign_taxonomy}.
Hardware exposes a security interface in the form of ISA extensions for supporting secure remote enclave execution.
Software is refactored to express its security requirements and map this requirements onto the software-exposed ISA extensions.
In the case of SGX, as a specific example, code is partitioned into enclave code and native code using SGX ISA extensions.
The result is confidentiality (with the exception of side-channels) for enclave data and integrity for enclave code and data.
In other words, hardware and software are working together to provide a particular user-specified level of application security.

\subsection{Homomorphic Encryption}

Homomorphic encryption (HE) is an emerging secure computing technology that enables computation \textit{directly over ciphertexts without decrypting the contents} and was proposed by Gentry~\cite{gentry09, gentry10}.
In non-HE encryption, a client encrypts the plaintext data and sends it to a cloud service provider; the cloud provider then decrypts the data back into plaintext where functions can readily be applied.
Once computation completes, the service provider encrypts the result and sends it back to the client.
In this setting an honest-but-curious adversary (cloud service provider), such as those assumed by~\cite{gazelle, cryptonets}, can see the plaintext copy of sensitive user data.
HE is different because it allows for a cloud service provider to directly perform computation on ciphertexts; as a result, in HE a cloud provider never sees the plaintext version of the data it is performing computation on.
This provides cryptographically strong guarantees against adversaries for the data transmitted to the datacenter for computation as only the client is able to decrypt the data.
HE provides security guarantees while still allowing an untrusted third party to perform useful computation on the encrypted data.

HE comes with a number of severe computational restrictions and overheads which mandate the need for software-hardware-security codesign considerations to be practical.
First, the performance overheads for executing homomorphically encrypted kernels are generally four to five orders of magnitude~\cite{gentry2011implementing}.
In other words, executing a single multiplication or addition in HE over two encoded ciphertexts is 10000$\times$ to 100000$\times$ slower than a single plaintext addition using a CPU ALU.
As a result, to bring HE computation back down to practical speeds, HE will require significant codesign, hardware acceleration, and optimizations to reduce this overhead as much as possible.
Recent work has shown that certain applications containing abundant application parallelism which can be aggressively exploited by the underlying specialized hardware accelerator are preferable~\cite{heax, cheetah}.

Second, encrypted ciphertexts in HE are each associated with a noise budget that degrades monotonically with successive computations.
If the noise budget is exceeded for a given ciphertext the decryption procedure will fail which yields a random result and effectively losing the encoded value.
The noise budget is governed by parameters associated with the HE encryption scheme which can be increased to increase the noise budget.
These encryption parameters also determine the size of the ciphertext representation.
Increasing the encryption parameter sizes to increase noise budget directly increases the compute datapath width and storage requirements for ciphertexts.
As a result, these encryption parameters must be carefully balanced to avoid wasting noise margins but still allow correct decryption.
Methods like bootstrapping can reduce the noise for an intermediary ciphertext but the process is impractically slow (an additional 4-6 orders of magnitude overhead)~\cite{sgx-he}.
Therefore, practical implementations of HE computations effectively have a limited logical depth for computation (i.e., \textit{leveled} HE) that they can support bound by the noise budget.

Third, HE is severely limited in the types of computations that it can support efficiently.
Most modern constructions of HE such as BFV~\cite{bfv}, BGV~\cite{bgv}, and CKKS~\cite{ckks} pack and encrypt multiple plaintext values into a single ciphertext representation.
Each value is packed into a slot in the ciphertext the same way values are packed into vector registers.
Furthermore, values are constrained to fixed point or integer which require quantizing the target application.
Modern HE schemes only support single instruction multiple data (SIMD) add, SIMD multiply,
and slot rotation, which allows swapping the values between slots in the vectors of the ciphertext representations,
over the ciphertext vectors,
Each of these three operations as well as the order in which they are executed increase the noise budget differently.
In theory, arbitrary computation can be constructed out of addition and multiplication operations but they are not efficient in HE as they require many logical layers of computation to implement.
This means that application selection is a key design consideration; vectorizable computation with fewer logical layers with minimal control logic is more efficient in HE.
For instance, statically schedulable computation such as matrix multiplication is more amenable to HE than computation that requires data dependent control flow like sparse matrix algebra.

Putting it all together, HE presents a number of software-hardware-security codesign opportunities summarized in \autoref{tab:codesign_taxonomy}.
The high overhead of HE mandates the need for aggressive hardware specialization but the maximum speed ups a hardware architecture can achieve is determined by available application parallelism.
Similarly, application selection determines the underlying computational behavior such as instruction mix, control flow, and amenability to vectorization which impacts the noise budget requirements to prevent decryption from failing.
This ultimately dictates the noise budget requirements and encryption parameter settings which affect the compute datapath width and storage requirements on the target hardware platform.
All together, we find that reaching optimized design solutions with HE is another example of an opportunity for software-hardware-security codesign.

\subsection{Other Secure Computing Technologies}

There are other instances (in varying stages of maturity) of secure computing technologies that benefit from hardware-software-security codesign insights; for instance, one group of secure computing technologies specifically addresses information leakage through side channels.
Consider systems where multiple tenants share some key computing resources such as caches.
This is the case in cloud computing applications, which store and compute over private data.
In cloud service infrastructures, virtual machines share the underlying hardware resources and isolation is usually not provided by default.
Stealthy cache attacks such as
  FLUSH$+$RELOAD~\cite{gruss2016flush, gullasch2011cache, yarom2014flush}, PRIME$+$PROBE~\cite{irazoqui2015s, kayaalp2016high, liu2015last, osvik2006cache, percival2005cache},
  EVICT$+$TIME~\cite{osvik2006cache} and
  others~\cite{disselkoen2017prime} are considered practical; this allows adversaries to steal secret AES~\cite{kayaalp2016high}, RSA~\cite{yarom2014flush}, and ElGamal~\cite{liu2015last} cryptographic keys, spy over encrypted channels~\cite{irazoqui2015lucky}, and log keys~\cite{gruss2015cache}.
A similar scenario also occurs in smart phones, where a malicious application can learn side-channel information about the system through shared resources such as caches~\cite{lipp2016armageddon}.

There are ways to mitigate against these side-channel attacks but identifying the root causes and mitigating them requires insights into both hardware and software behaviors to codesign them to meet security requirements.
For example, software running on vulnerable systems should be written in such a way that it avoids behavior that enables information leakage.
This can be done by writing code such that control flow or memory access patterns eliminate observable side effects that leak information through hardware-induced side channels.
Software countermeasures such as page coloring can also be used.
Other hardware solutions such as Intel's cache allocation technology, and attack resistant caches~\cite{liu2016newcache, wang2008novel, liu2014random} can mitigate these vulnerabilities but still require codesigning the underlying hardware systems to be aware of and properly mitigate such vulnerabilities.

The post-quantum cryptography competition~\cite{nist-quantum, nist-quantum2} illustrates another relevant opportunity for software-hardware-security codesign similar to the standardization of AES.
The primary motivation for the post-quantum cryptography effort stems from application concerns that quantum computing will defeat the hardness guarantees of existing cryptographic techniques, i.e., there is a need to re-codesign the software or encryption algorithm to strengthen security properties.
This illustrates the ongoing design feedback between security needs and the algorithms that realize them.
The latest round of the standardization effort focus on the security and cryptanalysis but, as with prior efforts such as AES, there is an opportunity for hardware and software codesign considerations to ensure that theoretically strong cryptography can be put into practice.

\section{Putting Systemization Into Practice}
\label{sec:themes}

Our systemization shows that to support the next generation of secure computing we must integrate knowledge across software, hardware, and security.
To do this, we need systematic rules of thumb to guide effective integration into the codesign process.
This section highlights several opportunities and focus areas to facilitate putting software-hardware-security codesign into practice.

\subsection{Security as a First-Order Design Principle}
Security is often treated as an afterthought in the software and hardware design process.
The Spectre~\cite{spectre} and Meltdown~\cite{meltdown} vulnerabilities exemplified the consequences of optimizing for performance first and fixing security issues in hardware afterwards.
The interconnected design considerations again make it more clear that security needs to become a first class constraint in design methodologies.
Understanding how software, hardware, and security interact is the first step towards unifying security to design methodologies.
This systemization provides the insights that can be translated into systematic rules of thumb to effectively guide the codesign process.
In other words, rules of thumb for security design considerations and their implications must be communicated to software and hardware experts that provide reasonable actionable steps when building a system.

There is already precedence in the hardware-software codesign space towards establishing systematic rules to guide codesign opportunities.
For instance, hardware designers often propagate design guidance to software designers that integer computation is preferred over floating point because it is more power and area efficient.
Thus, a good rule of thumb for software designers who are not aware of the underlying hardware implementation is to use integer math whenever possible.
Similar rules of thumb to guide hardware and software designers towards security-friendly designs are necessary to both establish this abstraction boundary and integrate security as a first class design consideration.

Our systemization study reveals that each technology needs to communicate unique design constraints and implementation challenges to the hardware and software.
For example, for TEEs and enclaves, a designer should minimize the amount of data transferred into and out of the enclave boundary in the application design to reduce the performance overhead.
For homomorphic encryption, an application designer should choose applications which are more amenable to vectorization and have limited control flow to minimize noise budget requirements.
For DP, selecting the right privacy budget to ensure that both application behaviors and any correlated information leakages at lower levels of the stack are sufficiently hidden.
Providing these succinct systematic guidelines between software, hardware, and security allows designers to translate the insights in one domain and effectively apply them to optimize design constraints in another.
More importantly, it provides understandable design rules of thumb to guide cross-stack optimization.

\subsection{Expose Usable Abstractions for Effective Codesign Interactions}

The second key theme that our study exposes is the trend towards interfacing primitives between software, hardware, and security.
In order to either automate or make the knowledge transfer between domains efficient, we need to provide a well-established vocabulary at the interface to collaborate effectively.
This is similar to how the instruction set architecture establishes a contract between what the software sees and what the hardware implements in modern architectures and accelerators (a hardware accelerator is essentially one large instruction).
The abstraction also serves as an interface between disciplines that allow experts in disparate fields to better communicate the challenges and systematic guidance towards co-optimizing solutions.

In the context of software-hardware-security codesign, a similar contract between security and software, and security and hardware remains an open question.
For some technologies, these abstraction boundaries and the primitives to leverage them are quickly maturing.
For instance, in DP, work in programming languages exposes a type system interface that guarantees DP of an application if properly leveraged and implemented against the API.
Work in verification and static analysis~\cite{near2019duet} augments this interface by providing a way for the application to verify that the DP security guarantees are properly implemented for the application.

Similarly, for secure enclave implementations like SGX, the security mechanism exposes an instruction set extension to provide the application developer an interface to leverage the security mechanisms provided by the enclave.
Homomorphic encryption also has a similar abstraction interface where a limited SIMD instruction set provides an interface for the application to reconcile constraints against the technology; there are several early proposals to provide compiler support to make the technology more accessible to programmers~\cite{eva, chet}.
These abstractions are all important because they help manage the complex design dependencies where designers can reason about the security guarantees.

\subsection{Uphold Abstraction Boundary Contracts}

Establishing clean abstraction boundaries and providing primitives to communicate across software, hardware, and security domains is useful to translate design-time requirements, but does not provide any guarantees on whether these requirements are fulfilled.
In other words, we need to verify that the contract between software and security or hardware and security is upheld.
For instance, a TEE is not useful if it provides an ISA but does not implement the hardware or software in a way that upholds the security guarantees that it claims.
As a result, it is also important to verify that the software, hardware, and security properties that are exchanged for complying with these abstraction interfaces are in fact implemented correctly.

For instance, an application can receive confidentiality guarantees (for data) and integrity guarantees (for code and data) via hardware support for properly partitioned programs in TEEs.
Here, software is responsible for providing a partitioning that properly leverages the TEE interface.
In the case of enclaves like SGX, contract between software, hardware, and security is verified via remote attestation in which the software verifies that the software verifies that the hardware system that is offering the secure execution guarantees is genuine.
The contract that the hardware implementation of the enclave is correct and implemented correctly is rooted in the manufacturer (i.e., security to hardware contract).

A similar set of contracts appears in the design considerations for homomorphic encryption.
In HE, the software interfaces with security through noise budget constraints and encryption parameters.
If the software implementation satisfies the noise budget constraints using the provided encryption parameters, HE guarantees computing a correct result; a violation of this contract (i.e., exceeding noise budget) results in the decryption failure scenario which is can be verified with testing.
In HE, an application also must conform to the restricted HE instruction set which is a canonical software-hardware contract but in this case the contract also includes restrictions from security.

Moving forward, it will be important to establish both the abstraction boundaries and verifiable guarantees.
Threat models and root of trust definitions provide these interfaces to some degree but still are missing the verification aspects.
Threat models provide the desired security requirements via adversary capability definitions but do not specifically address mechanisms for how threats are verified to be mitigated by software or hardware.
Many of these abstractions boundaries and contracts remain an open question as to what the correct verifiable manifestations is.
For instance, it is not clear what a contract may look like between DP and the hardware implementation of a DP software.
Establishing ways to verify these contracts thus remains an area of future work that can bolster the design of private and secure systems.

\section{Looking Forward}

\label{sec:future-work}

Our systemization illustrates that the codesign process for secure computing technologies is an interdisciplinary process which currently requires deep specialized knowledge across applications, hardware, and security to conduct.
In other words, only an expert or group of experts who understand all three disciplines can successfully perform the codesign exercise.
Going forward, to facilitate wider adoption and deployment, experts who understand the software-hardware-security codesign considerations will need to build automation support and infrastructure to abstract away the expertise requirements so that non-experts can focus on building on top of these technologies.
To do this, we need to build accessible and robust tooling infrastructure similar to how LLVM and GCC compiler stacks abstracted away the complexities of compilation passes and optimization to allow programmers to focus on building applications.

Ongoing efforts towards these types of tools include compilation support for homomorphic encryption~\cite{chet, eva}, automatic partitioning for SGX~\cite{lind2017glamdring}, and programming abstractions for differential privacy~\cite{barthe2016programming, mcsherry2009privacy, roy2010airavat}.
In each case, the tools abstract away the details of how the underlying secure computing technology is implemented and automates the specialized expertise so that it can be reused by non-experts.
This allows a designer who may have no knowledge of these secure computing technologies to still leverage the technology.
While many efforts in this space are still nascent, these are the sorts of tools that will enable wider adoption by designers and put secure computing technologies into practice.

\section{Conclusion}

We present a systemization of knowledge for secure computing technologies and highlight opportunities for cross-stack software-hardware-security codesign.
For each technology, we identify the codesign considerations across software, hardware, and security, and expose the design feedback considerations that mandate co-optimization across each to guide efficient system design and integration.
Insights into both emerging technologies as well as historical precedence can both guide and motivate the same codesign opportunities for future technologies.
By recognizing the need for co-optimizing secure computing solutions in both historical and modern secure computing techniques, we hope these will inform more systematic and efficient deployments in the future.



\bibliographystyle{ACM-Reference-Format}
\bibliography{references}


\end{document}